\documentclass[a4paper,12pt,fleqn]{article}
\usepackage{amsmath,graphicx}

\newcommand{\pni}{\par\noindent}

\begin{document}
\title{Black hole horizon and space-time foam}
\author{ A. G. Agnese\footnote{Email: agnese@ge.infn.it} \, and
M. La Camera\footnote{Email: lacamera@ge.infn.it}} 
\date{}
\maketitle 
\begin{center}
\emph {Dipartimento di Fisica dell'Universit\`a di 
Genova\\Istituto Nazionale di Fisica Nucleare,Sezione di 
Genova\\Via Dodecaneso 33, 16146 Genova, Italy}\\
\end{center} 
\bigskip
\begin{abstract}
We introduce, by means of the Brans-Dicke scalar field, 
space-time fluctuations at scale comparable to Planck length 
near the event horizon of a black hole and examine their dramatic
effects. 
\end{abstract} 
\bigskip\bigskip\bigskip \pni
PACS numbers: \; 04.20.Cv , 04.50.+h , 04.60.-m
\vspace{1in}\pni
\newpage
\baselineskip = 2\baselineskip

Is is  generally accepted  that classical description of 
space-time breaks down to ``quantum foam'' [1] when the spatial 
distance between two points is of the order of Planck length 
$l_P$, the  underlying idea being that space-time topology 
fluctuates at those distances.  \par 
Moreover in recent papers Amelino-Camelia [2] and Ng and 
van Dam [3], basing on a gedanken timing experiment 
originally devised by Salecker and Wigner [4], argue that such
a distance could be much greater that Planck length. As a 
consequence quantum gravity effects could be probed with
current or future interferometers designed for the 
gravitational-wave detectors.\par In this letter, starting from a
different point of view, we give some arguments which can lead, 
already when distances vary at the Planck scale, to dramatic 
effects near the event horizon of the Schwarzschild vacuum 
solution now perturbed by the corresponding fluctuations of 
space-time. We will show that the metric breaks down, even for 
large masses, in correspondence of the surface which should have 
been the event horizon, so it becomes unpossible not only to 
define an event horizon but even to foresee which singularity 
has replaced it. These facts signal the need 
of a fully quantum description of gravity in the presence of 
distances suitable for the Planck regime even if surfaces are 
much greater than $l_P^2$ and the gravitational field in their 
neighbourhood is weak.\par To achieve this goal we start by 
considering the static spherically symmetric vacuum solution of 
the Brans-Dicke theory of gravitation [5]. \par The related 
calculations were performed by us in Ref. [6], working in the 
Jordan frame, where the action is given (in units $G_0 = c = 1$) 
by \begin{equation} S = \dfrac{1}{16 \pi }\,\int\, d^4x \sqrt{-\,
g}\left[ \Phi R - \, \dfrac{\omega}{\Phi}\nabla^\alpha \Phi 
\nabla_\alpha \Phi \right] \end{equation}
and with a suitable choice of gauge. Here we quote only the 
results relevant for the following.\par
The line element can be written as
\begin{equation}
ds^2 = e^{\mu(r)} dr^2 +  R^2(r) d\Omega^2 - e^{\nu(r)} dt^2
\end{equation}
where $d\Omega^2 = d\vartheta^2 + \sin^2 \vartheta d\varphi^2$ 
and, in the selected gauge:
\begin{align*}
R^2(r) = r^2 \left[1-\dfrac{2\eta}{r}\right]^{1-\gamma 
\sqrt{2/(1+\gamma)}} \tag{3a} \\ {} \\
e^{\mu(r)} = \left[ 1 - \dfrac{2\eta}{r}\right]^{-\gamma 
\sqrt{2/(1+\gamma)}}  \tag{3b} \\ {} \\
e^{\nu(r)} = \left[1 - 
\dfrac{2\eta}{r}\right]^{\sqrt{2/(1+\gamma)}} \tag{3c}
\end{align*}
\setcounter{equation}{3}
Here $\gamma$ is the post-Newtonian parameter 
\begin{equation}
\gamma = \dfrac{1+\omega}{2+\omega}
\end{equation}
and
\begin{equation}
\eta = M \sqrt{\frac{1+\gamma}{2}}
\end{equation}
Finally  the scalar field is given by
\begin{equation}
\Phi(r) = \Phi_0 \left[1-\dfrac{2\eta}{r}\right]^{(\gamma 
-1)/\sqrt{2(1+\gamma)}}
\end{equation}
while the effective gravitational coupling  $G(r)$  equals
\begin{equation}
G(r) = \dfrac{1}{\Phi(r)}\,\dfrac{2}{(1+\gamma)} 
\end{equation}
the factor $2/(1+\gamma)$ being absorbed, as in Ref. [5], in the 
definition of $G$.\par
Departures from Einstein's theory of General Relativity appear 
only if $\gamma \neq 1$, a possibility consistent with 
experimental observations  which estimate it in the range \, 
$1-0.0003 < \gamma < 1+0.0003$ corresponding to the dimensionless
Dicke coupling constant $|\omega|> 3000$.\par When $\gamma <1$ 
and $r \to 2\eta$, then $R(r)$, $e^{\nu(r)}$ and $G(r)$ go all 
to zero. Therefore we have a singularity with infinite red-shift 
and gravitational interaction decreasing while approaching the 
singularity.\par When $\gamma > 1$, the null energy condition 
(NEC) is violated [8] and a wormhole solution is obtained [6] 
with throat at 
\begin{equation}
r_0 = \eta \left[1+\gamma\,\sqrt{\dfrac{2}{1+\gamma}}\,\right]
= M \left[ \gamma + \sqrt{\dfrac{1+\gamma}{2}}\right]
\end{equation}
to which corresponds the value $R_0$ given by equation (3a). It 
is easy to verify that $r_0 > 2\eta$ and $R_0 > 0$. Now the 
singularity is beyond the throat and is smeared on a spherical 
surface, asymptotically large but not asymptotically flat, of 
radius $R(r) \to \infty$  as $r \to 2\eta$ and where also the 
red-shift and $G(r)$ become infinitely large [8,9].\par
When $\gamma = 1$ exactly, one has Schwarzschild solution of 
General Relativity
\begin{equation}
ds^2 = \dfrac{dr^2}{1-\dfrac{2M}{r}}+r^2 d\Omega^2- (1-
\dfrac{2M}{r}) dt^2
\end{equation}
where it appears a black-hole with event horizon at $R(r)=r 
=2M$. \par
If future measurements will establish that $\gamma \neq 1$ and 
if the Jordan frame is the physical frame, then the strong 
equivalence principle is violated (due to exotic matter in the 
case $\gamma < 1$) and it will possible to decide on the type of 
singularity  occurring. Moreover gravitation 
shall be better described by a suitable generalization of 
Einstein's theory. \par
Here we first assume $\gamma = 1$ exactly but then we introduce 
a fluctuation in the Schwarzschild metric by imposing a violation
of the strong equivalence principle, of the order of the Planck 
length $l_P$, which comes from the following variation of the 
gravitational radius : 
\begin{equation}
2\eta = 2M  \pm \dfrac{l_P}{2}
\end{equation}
Making use of Equation (5) one obtains for $\gamma$, at first 
order in $l_P/M$
\begin{equation}
\gamma = 1 \pm \frac{l_P}{M}
\end{equation}
so from Equation (4) the dimensionless coupling constant 
$\omega$  turns out to be $|\omega| \approx M/l_P$.
To make an example, in the case of a solar mass $M_{\bigodot}$ 
this would amount to a fluctuation of
$\Delta M_{\bigodot}/M_{\bigodot} \approx 10^{-38}$.\par
It may be useful to rewrite, in this approximation, the metric 
coefficients of the Brans-Dicke line element:
\begin{align*}
R^2(r) = r^2 \left[1-\dfrac {2 M \pm 
\dfrac{l_P}{2}}{r}\right]^{\mp \frac{3 l_p}{4 M}} \tag{12a}\\{}\\
e^{\mu(r)} =  \left[1-\dfrac {2 M \pm 
\dfrac{l_P}{2}}{r}\right]^{-1 \mp \frac{3l_P}{4 M}} 
\tag{12b}\\{}\\
e^{\nu(r)} =  \left[1-\dfrac {2 M \pm 
\dfrac{l_P}{2}}{r}\right]^{1\mp \frac{l_P}{4 M}} \tag{12c} 
\end{align*} 
\setcounter{equation}{12}
Here the upper sign refers to the case $\gamma >1$ and the lower 
one to the case $\gamma <1$. It is apparent that, while in the 
case $\gamma = 1$  the event horizon is fixed at $r_S = 2M$, in 
the presence of space-time fluctuations one may have, taking the 
extreme values of $\gamma$ given by Equation (11), either a naked
singularity at $r_< = 2M - l_P/2$ if $\gamma = 1 - l_P/M$ or a 
wormhole with a throat at $r_0 = 2M + (5/4)\, l_P$ if $\gamma = 1
+ l_P/M$; in this latter case we have beyond the throat at $r_> =
2M + l_P/2$ another singularity which is asymptotically large but
not asymptotically flat. We recall that $r_<$ and $r_>$ are the 
limits between which we required to fluctuate the metric. 
When $r$ exceeds $2M$ by several units of $l_P$ the 
solution of the field equations behaves as in the Schwarzschild 
case, but when $r$ varies around $2M$ in an interval 
containing all the possible singularities one is faced with 
different interchanging behaviours which are unpredictable on 
classical grounds. In this latter case the proper distance from 
the singularity of a point near to $2M$ is proportional to 
$\sqrt{M l_P}$. To make a numerical example in the case $\gamma 
>1$, the proper radial distance of $r_>$ (to which corresponds 
an infinite value of the standard radial coordinate $R$) from 
$r_0$ (to which corresponds the location $R_0$ of the throat) is
\begin{equation} 
L = \int_{r_>}^{r_0} e^{\mu(r)/2}dr \approx 
2.51 \sqrt{M l_P}
\end{equation}
and the proper volume comprised between the radii is
\begin{equation}
V = 4\pi \int_{r_>}^{r_0} R^2(r) e^{\mu(r)/2}dr \approx
36 \pi \sqrt{M^5 l_P} 
\end{equation}
For the Sun the numerical values are $L = 3.9\,10^{-16}\, m$ and 
$V = 3.4\, 10^{-10}\, m^3$. So, if it is  correct to
introduce  space-time fluctuations by means of the 
Brans-Dicke scalar field, then it would be needed a full quantum 
theory of gravitation to describe not only the kind and the 
evolution of the singularities but also the space-time that 
surrounds them at distances much greater than the  Planck 
length. If the scenario we have proposed here is realistic, 
it should turn out to be a good candidate both as a source of 
strong gravity waves and as a central engine required for the 
production of gamma ray bursts. 

\end{document}